\documentclass[reprint,
superscriptaddress,
amsmath,amssymb,
aip,
]{revtex4-1}

\usepackage{graphicx}
\usepackage{dcolumn}
\usepackage{bm}
\usepackage{amsmath}
\usepackage{color}
\usepackage{float}
\usepackage[hidelinks]{hyperref}
\usepackage{subcaption}
\usepackage[section]{placeins}

\graphicspath{{./Figures/}}  

\newcommand{\prlsec}[1]{\textit{#1.---}}

\begin{document}

\pretolerance=10000
\tolerance=2000
\emergencystretch=10pt

\displaywidowpenalty = 10000

\title{Characterization of Non-linearities through Mechanical Squeezing in Levitated Optomechanics}

\author{Ashley Setter}
\email{A.Setter@soton.ac.uk}
\affiliation{%
 Department of Physics and Astronomy, University of Southampton, SO17 1BJ, United Kingdom 
}%

\author{Jamie Vovrosh}
\altaffiliation[Present affiliation: ]{%
  School of Physics and Astronomy, University of Birmingham, Birmingham, B15 2TT, United Kingdom 
}%
\affiliation{%
 Department of Physics and Astronomy, University of Southampton, SO17 1BJ, United Kingdom 
}%

\author{Hendrik Ulbricht}%
 \email{H.Ulbricht@soton.ac.uk} 
\affiliation{%
 Department of Physics and Astronomy, University of Southampton, SO17 1BJ, United Kingdom 
}%

\begin{abstract}
  We demonstrate a technique to estimate the strength of non-linearities present in the trapping potential of an optically levitated nanoparticle. By applying a brief pulsed reduction in trapping laser power of the system such as to squeeze the phase space distribution and then matching the time evolution of the shape of the phase space distribution to that of numerical simulations, one can estimate the strength of the non-linearity present in the system. We apply this technique to estimate the strength of the Duffing non-linearity present in the optical trapping potential.
\end{abstract}

\maketitle

Non-linearities have been proposed to be used in optomechanical systems for inducing steady-state mechanical squeezing \cite{Lu2015}, ground state cooling \cite{Momeni2018}, achieving sub-Poissonian statistics \cite{Grimm2016} and generating non-Gaussian states \cite{Aspelmeyer2014}. It has been demonstrated that mechanical non-linearities can be mapped into the microwave domain by coupling with a Microwave superconducting coplanar waveguide (CPW) resonator \cite{Zhou2013} and have been exploited to improve the figures of merit for nanotube and graphene resonators \cite{Eichler2011}.

For cantilever systems it has been proposed that one could use a Duffing non-linearity to observe classical to quantum transitions utilizing the bistable regime of such an oscillator \cite{Katz2007}. Non-linearities have been utilized alongside frequency stabilization to achieve large amplitude and therefore large signal-to-noise ratio in MEMS devices \cite{Antonio2012}, to realize mechanical bit operations \cite{Badzey2004, Mahboob2008} and to induce stochastic switching to boost detected signals \cite{Venstra2013}.

Non-linearities can also be exploited to differentiate between classical and quantum dynamics; as in a harmonic potential such behaviour is difficult to distinguish \cite{Katz2007, Ralph2018}. Precise control of the non-linear and stochastic bistable dynamics has been achieved for optically levitated nanoparticles within a Duffing potential and used to demonstrate stochastic resonance in good agreement with analytical models and utilized to amplify forces on the system \cite{Ricci2017}.

The method we describe here can be used to determine non-linearities which are inherent to the system \cite{Gieseler2013a}, as well as those introduced by external forces \cite{Diehl2018,Winstone2018} which perturb the potential. Estimating non-linearities of optomechanical systems is an area of interest and several methods have been proposed and utilized to probe non-linearities \cite{Gieseler2013a, Latmiral2016, Ricci2017}.

In this paper we demonstrate characterization of the inherant Duffing non-linearity in the mechanical potential by applying an operation to the system where the power of the trapping laser is reduced rapidly for a brief pulse before being restored to the original power in such a way as to sqeeze the phase space distribution of the system. We perform many such pulses and then match the resulting phase space distribution to numerical simulation in order to estimate the strength of the Duffing non-linearity. This method can be applied to any physical system where the phase space distribution can be sqeezed and subsequently allowed to freely evolve in a region of phase space where non-linearities of the system affect the evolution. In this paper we find that a perturbation of the potential by a factor as low as $\frac{V(z)}{25}$ is sufficient for a visible effect on the phase space distribution.

\begin{figure}[t!]
  \centering
  \includegraphics[width=0.45\textwidth]{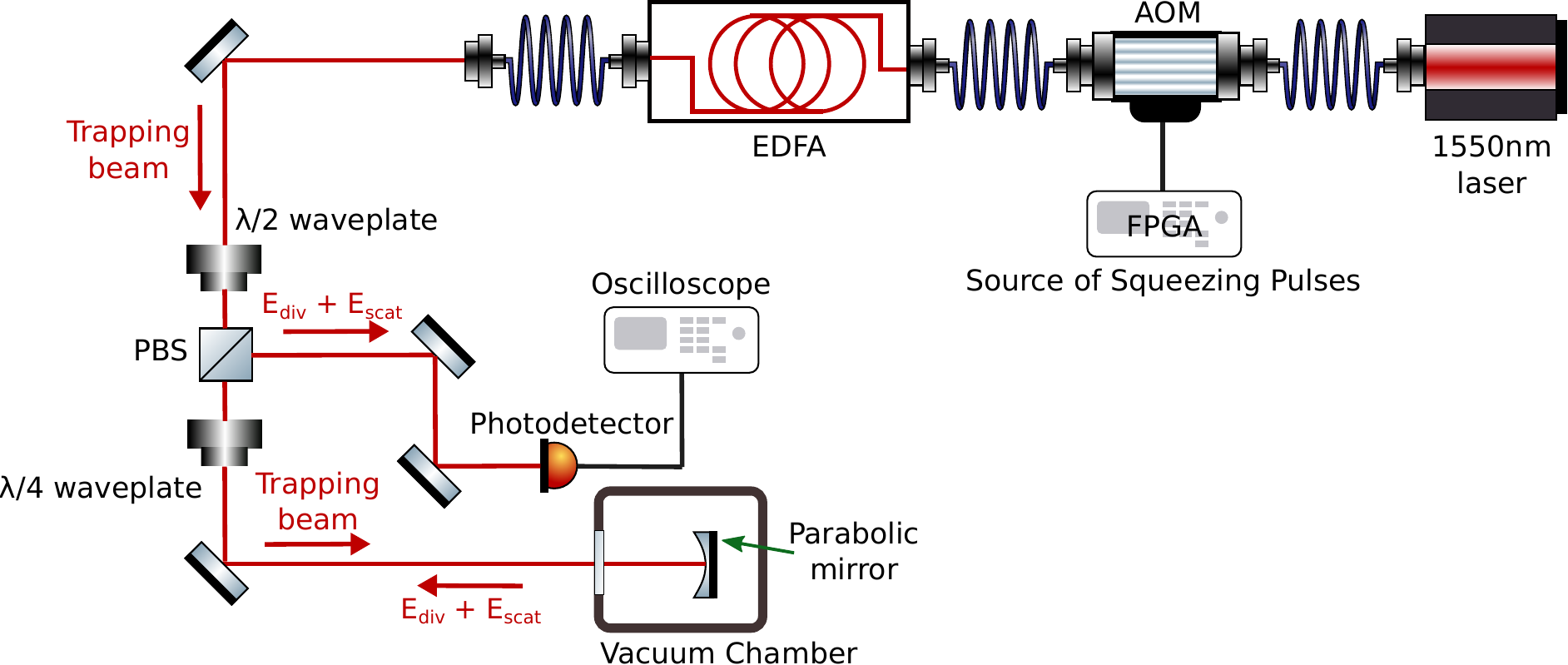}
  \caption{\label{ExperimentalSetup}
    The 3D position of the particle is detected by interference of the scattered and divergent field by the photo-detector. The squeezing pulse operation is fed to an AOM to rapidly switch the power of the seed laser. The Erbium Doped Fibre Amplifier (EDFA) then amplifies this modulated light to the power required to trap the nanoparticle.
  }
\end{figure}

The experiment was performed using the setup shown in figure \ref{ExperimentalSetup}. To perform a squeezing operation we first rapidly change the natural oscillation frequency from $\omega_0$ to $\omega_1$ then we let the system evolve for time $\tau_{Pulse} = \dfrac{\pi}{2\omega_1}$ before rapidly switching back to $\omega_0$ \cite{Rashid}. Such a frequency modulation is achieved using an acousto-optic modulator (AOM) to rapidly reduce the trapping laser power from $P_0$ to $P_1$ and, after a short delay, raise the power back to $P_0$.

\begin{figure}[t!]
  \centering
  \includegraphics[width=0.45\textwidth]{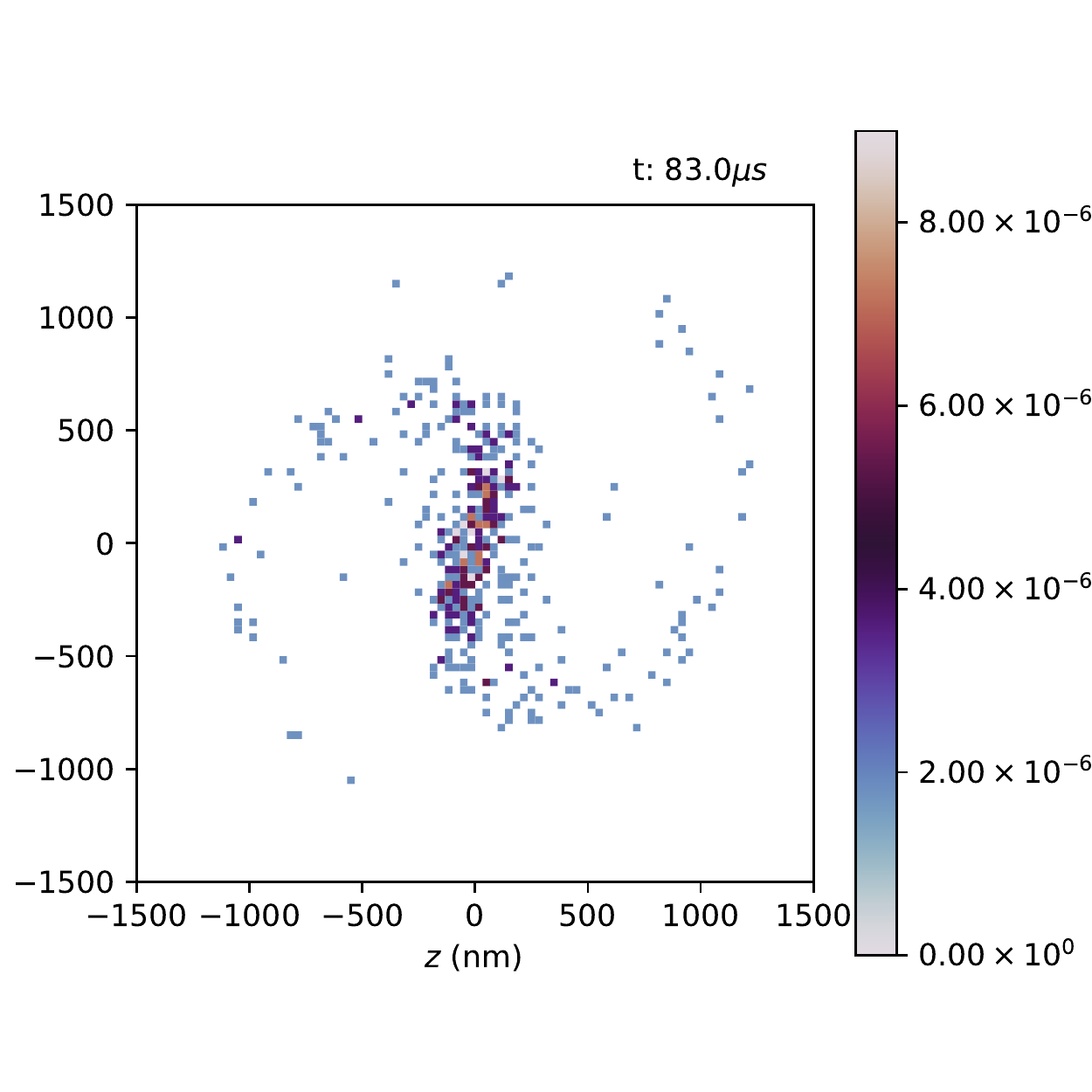}
  \caption{\label{ExperimentalPhaseSpace}
    The extracted points in phase space, accumulated over 500 experimental runs, $83\,\mu s$ after the squeezing pulse is applied.
  }
\end{figure}

In this particular experiment the translational motion parallel to the laser propagation direction was used, which had a natural frequency of $64.9\pm0.3\,kHz$ and the power was decreased by $78.4\pm1.5\,\%$ for $8.29\,\mu s$, which corresponds to a quarter of the natural frequency at the lower laser power during the pulse, which was $30.16\pm0.45\,kHz$. The squeezing pulse was applied and the then system allowed to relax while being measured.  We performed this operation 500 times, measuring the photodetector current for each pulse and relaxation. If we then plot the ensemble of the points of position and velocity, accrued over the 500 operations, $83\,\mu s$ after the pulse, we build up the phase space distribution shown in figure \ref{ExperimentalPhaseSpace}. These experiments were performed at a pressure of $0.164 \pm 0.025 mbar$.

The spiral shape we observe in the phase space in figure \ref{ExperimentalPhaseSpace} can be explained by the optical potential taking the form of a softening Duffing potential. The points where the particle is displaced further from the centre of the potential well experience a softer potential well with a lower frequency. Therefore, points further from the origin in phase space lag behind the more central points with a lag proportional to their displacement, resulting in the spiral pattern we experimentally observe in phase space.

The second order correction to a harmonic potential for a focused Gaussian beam is a Duffing non-linearity such that the potential takes the form $V(q) = \omega_0^2 q^2 + \omega_0^2 \xi q^4 = \omega_0^2(1 + \xi q^2)q^2$, where $\omega_0$ is the frequency of the harmonic component of the potential, $\xi$ is the Duffing non-linearity and $q$ is the displacement.

The motion of the nanoparticle in this potential is well described by the following classical stochastic differential equation \cite{Gieseler2013a},

\begin{equation}
  \ddot{q} = -\Gamma_0 \dot{q} + [1 + S_q](- \omega_0^2 q - \xi \omega_0^2 q^3) + \sqrt{\dfrac{2 \Gamma_0 k_B T_0}{m}} \dfrac{dW}{dt}
  \label{sde}
\end{equation}

where $\Gamma_0$ is the damping on the system due to gas collisions, $W$ is a real, zero-mean, Wiener process, $T_0$ is the temperature of the surrounding gas environment and $m$ is the mass of the nanoparticle. $S_q$ is a term parameterizing the squeezing pulse and takes the following form:

\begin{equation}
  S_{q} =
  \begin{cases}
    0.784, &  \text{if } 0\leq t\leq T_{pulse},\\
    0, & \text{otherwise}\\
  \end{cases}
\end{equation}

This equation neglects the effects of photon recoil because the effect is negligible at the pressure range considered here\cite{Jain}. 

From the experimental data we can extract the values of $\omega_0=408\times10^3\pm4\times10^3\,s^{-1}$ and $\Gamma_0=619\pm93\,\text{Hz}$\, \cite{Vovrosh2017} and we experimentally control the values of $S_q$ which are applied to the system. We also assume $T_0$ to be at room temperature, $300\,K$, as the gas inside the vacuum chamber is assumed to be in equilibrium with the environment. The range of $m$ allowed was narrowed down by analyzing the data \cite{Vovrosh2017} and then estimated more precisely to have a value of $m=4.8\times 10^{-19}\,kg$ by matching to simulation. This can be done because the extent to which the phase space distribution is squeezed is affected by the mass, but the shape of the spiral is not, so with a higher mass one gets the same shape in phase space but the arms of the spiral are less densely populated and more clumped in the centre.

This leaves $\xi$ as the only free parameter to vary in order to match the simulated phase space with the experimental phase space. The tightness of winding of the arms in the phase space distribution is determined by the effective frequency shift in the motion due to the strength of the Duffing term in the potential. As such the spiral shape in phase space is determined entirely by the magnitude of the Duffing term.

\begin{figure}
  \centering
\subcaptionbox{phase space after $83\,\mu s$ for $\xi = 0.0\,\mu m^{-2}$ in grey, $\xi = 0.05\,\mu m^{-2}$ in yellow, $\xi = 0.1\,\mu m^{-2}$ in green and $\xi = 0.2\,\mu m^{-2}$ in purple. \label{fig3:a}}{\includegraphics[width=.38\textwidth]{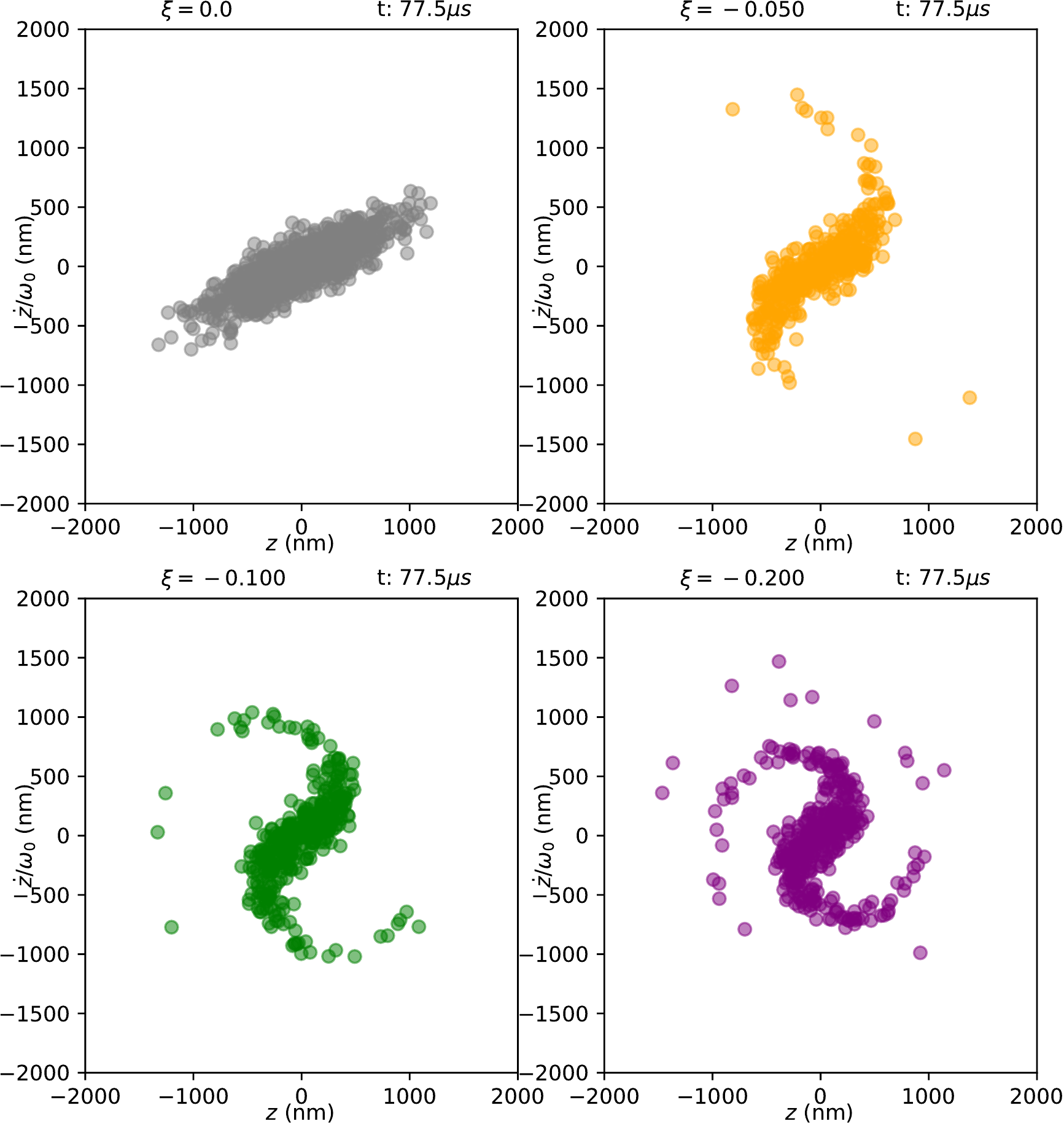}}\hspace{1em}%
  \subcaptionbox{Plot of the potential for the four values of $\xi$ with the colors corresponding to \ref{fig3:a} with $\xi = 0.0$ showing the harmonic component of the potential with no non-linearity present.\label{fig3:d}}{\includegraphics[width=.30\textwidth]{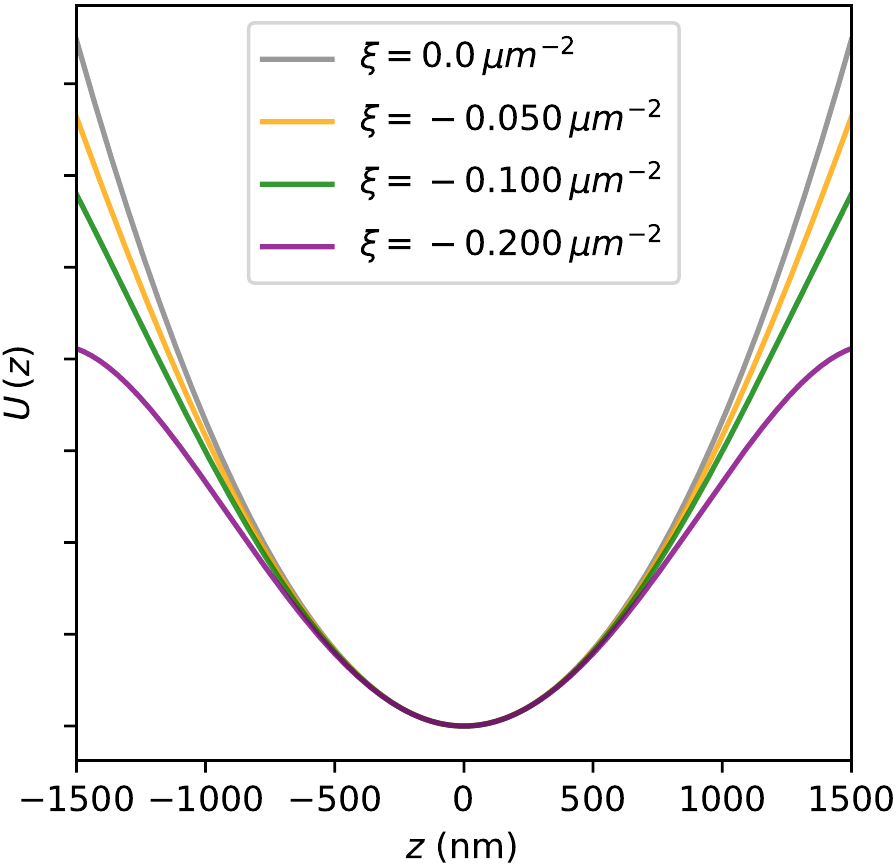}}
\caption{plots of phase space distributions $77.5\,\mu s$ after the squeezing pulse with four different $\xi$ values as well the potentials.}
\label{multiple_xi_vals}
\end{figure}

The phase space distribution accumulated over $400$ simulated trajectories is shown in figure \ref{multiple_xi_vals} for four values of $\xi$, demonstrating the relationship between $\xi$ and the tightness of the spiral. As the non-linearity strength, $\xi$, increases so does the gradient with which the frequency of the motion changes with position. In addition the trapping potential is plotted for the four values of $\xi$.

\begin{figure}[t!]
  \centering
  \includegraphics[width=0.45\textwidth]{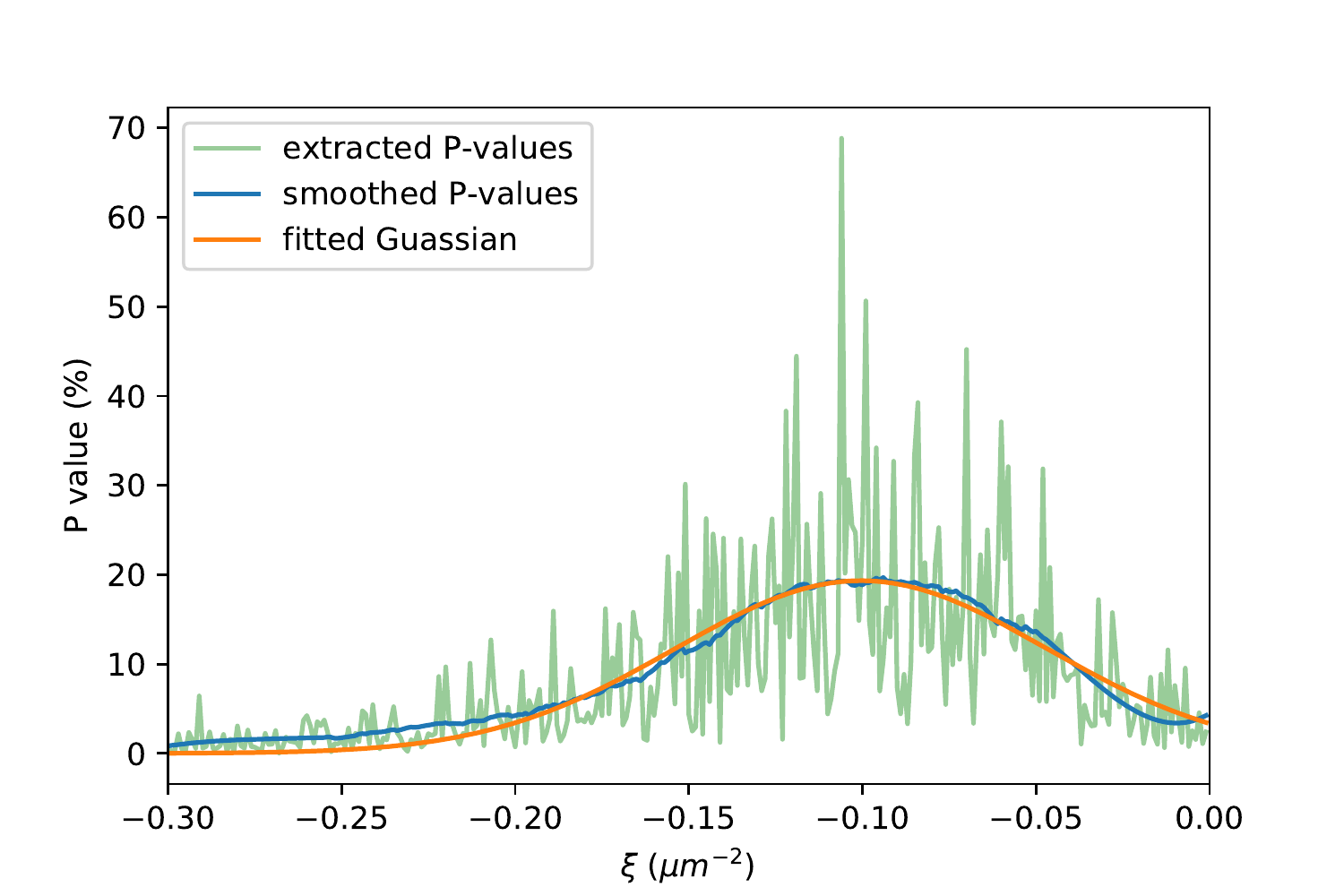}
  \caption{\label{P_values_with_xi}
    Plot of the P-values extracted by performing a Kolmogorov-Smirnov statistical test on 500 experimental data points and 400 simulated data points $77.5\mu s$ after the squeezing pulse for a range of 300 values of $\xi$ along with the same data with a smoothing Savitzky-Golay filter applied and the Guassian one obtains when fitting to this data to extract the stastistical mean and standard deviation. 
  }
\end{figure}

By taking the experimental and simulated phase space points a certain time after the pulse has finished one can compare their distributions using a Kolmogorov-Smirnov statistical test. This test returns a P-value on the null hypothesis that the two sets of points are drawn from the same distribution. This P value can be maximized to find the Duffing parameter $\xi$ for which the simulated phase space distribution best matches the experimental distribution. In order to find the value of $\xi$ for which the simulation best matches the experimental data we ran 400 simulations for 300 different values of $\xi$ and extracted the P-value for each, as can see seen in figure \ref{P_values_with_xi}. We then applied smoothing and fitted a Guassian to this data in order to extract the stastistical mean of $\xi=-0.100\,\mu m ^{-2}$ with a standard devation of $\sigma_{\xi}=0.054\,\mu m ^{-2}$.

\begin{figure}[t!]
  \centering
  \includegraphics[width=0.45\textwidth]{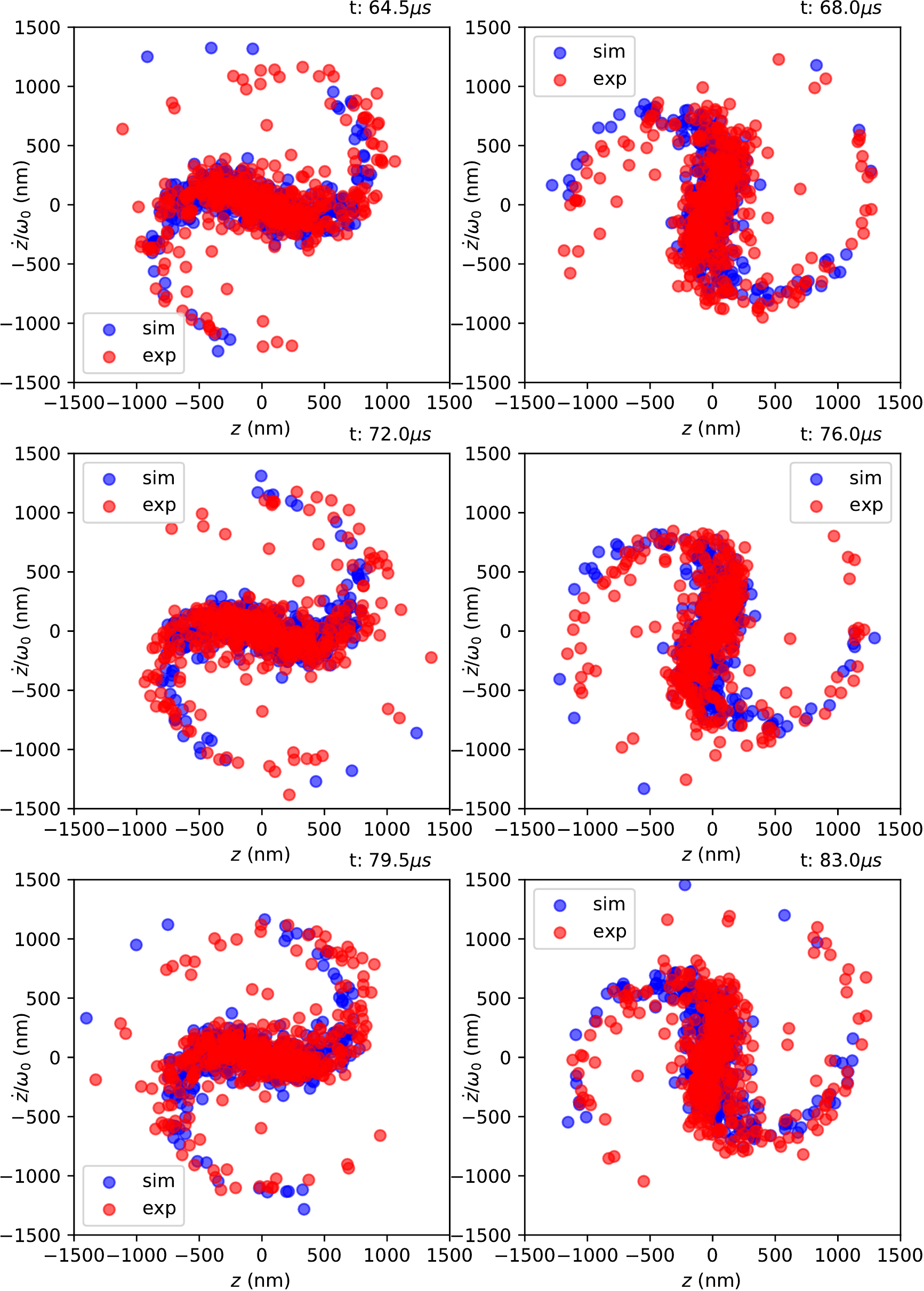}
  \caption{\label{MatchingPhaseSpacePlots}
    The extracted points in phase space, accumulated over 500 experimental runs, in red, and 400 simulated trajectories, in blue, at a number of sequential time instances, $t$, after the squeezing pulse operation is applied.
  }
\end{figure}

The meaning of the negative sign on this value is that as displacement from the centre of the trap gets larger the natural frequency of the motion gets lower, this is because of the Gaussian profile of the laser beam, as you get further from the centre of the trap the gradient of the electric field decreases non-linearly. With the stastistical mean value of $\xi=-0.100\,\mu m ^{-2}$ for the Duffing non-linearity, the experimental and simulated phase space distributions are plotted for a number of time instances after the pulse in figure \ref{MatchingPhaseSpacePlots}.

For the non-linear component of the potential to dominate over the harmonic component at a particular displacement $q$ one requires that $\xi > \frac{1}{q^2}$. If one takes for example a displacement of $500\,\text{nm}$, where in the phase space distribution the non-linear component can be seen to begin having a visible effect, one calculates that a $\xi$ of $4\,\mu m ^{-2}$ is the critical value above which the non-linearity would dominate the potential. We remark that the $\xi$ value we extract is much lower than this, meaning a dominant non-linear effect is not nessesary for the non-linearity to have a significant effect on the dynamics of the phase space distribution's evolution and for the method we have described here to be applicable.

\begin{figure}[t!]
  \centering
  \includegraphics[width=0.45\textwidth]{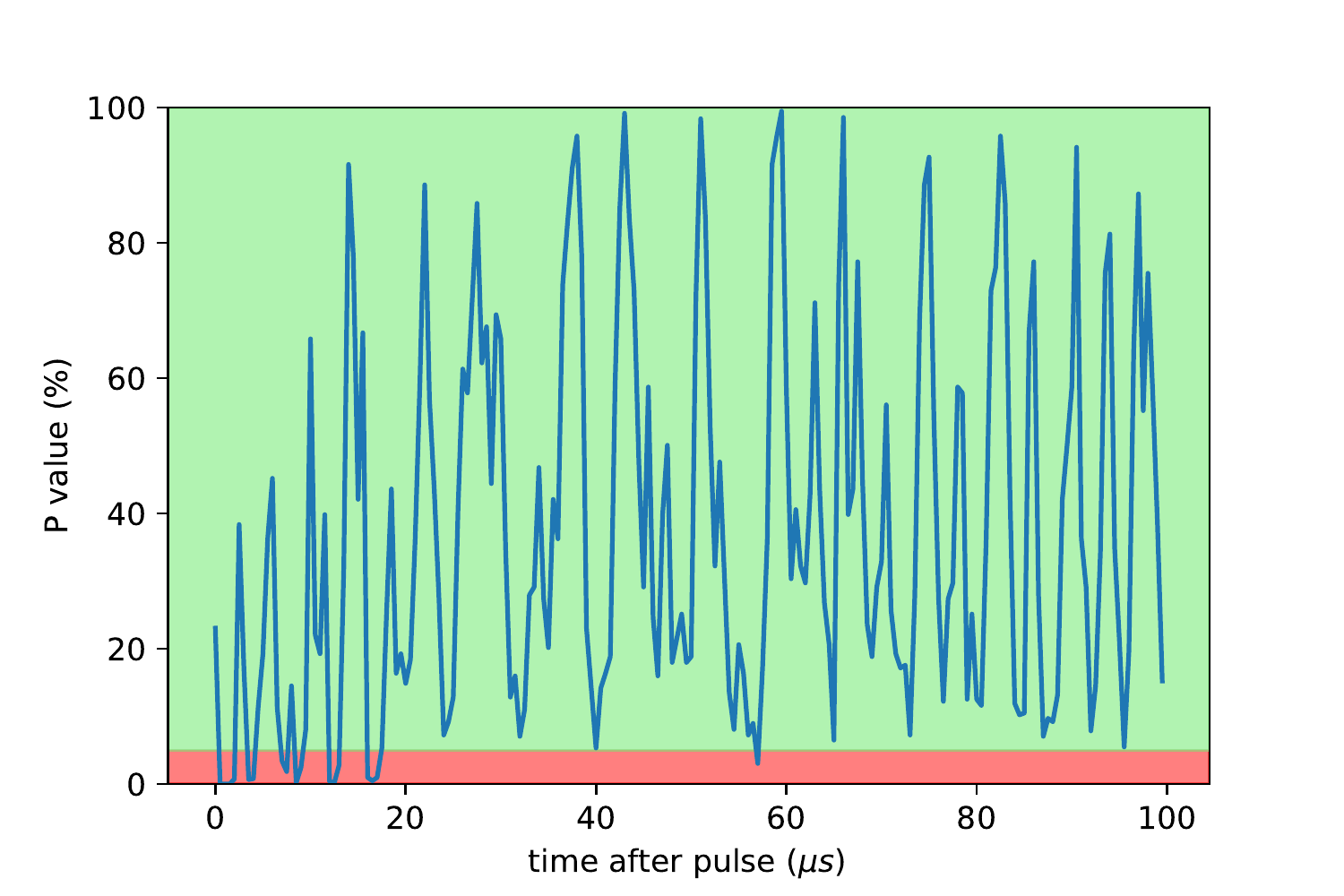}
  \caption{\label{Pvalue_timeplot}
    A plot of the P values with time for the null hypothesis that the two sets of points, the experimental phase space points and the numerically simulated phase space points, are drawn from the same distribution. The green region is where P-values are above the significance level $\alpha=0.05$ and red region is where the P-values are below this significance level. These P values are calculated using the Kolmogorov-Smirnov test applied to the collated experimental and simulated phase space data.
  }
\end{figure}

For a chosen $\xi$ value the P value can be plotted over time after the pulse, for the $\xi$ we find to be the best fit, $\xi=-0.100\,\mu m ^{-2}$, we obtain the plot of the P-value over time shown in figure \ref{Pvalue_timeplot}. From this figure one can observe that the P-value is distributed mainly above the 5\% significance level. It can also be observed that the P-value grows over time after the pulse. The reason for this is that the experimental phase space distribution is extracted by performing a band-pass filter upon the photo-current data and this processing distorts the initial position and velocity data due to edge effects from the filter.

In conclusion we have performed squeezing pulses on the system to observe the spiral shape in the phase space distribution resulting from a Duffing non-linearity. We have then simulated the trajectory of motion of the system by numerically solving the stochastic differential equations modelling the system. By varying the unknown Duffing strength and matching the simulated phase space distribution to the experimental one we estimate the Duffing strength.

The advantages of this method are that it is repeatable and can be performed on command and it does not rely on stochastic excitations in order to explore the non-linear regime. It can also be performed at pressures from $1\times10^{-1}-\times10^{-9}\,\text{mbar}$ as it does not rely on gas collisions to drive the system into the non-linear regime and only requires $Q$ factors high enough to observe the time evolution of the spiral shape for several oscillations. If performed at low pressures the high Q factor of the oscillator could in fact allow for more precise estimation of the non-linearity, since the spiral pattern will persist in the phase space for a longer time. In addition, feedback cooling can be applied while performing this procedure if the effect of the non-linearities are discernable at smaller displacements and if there is a risk of the particle escaping the trap\cite{Vovrosh2018}. 

This method to estimate non-linearities is also broadly applicable to a large range of physical systems such as levitated systems \cite{Pedregosa-Gutierrez2010, Gieseler2013a}, optomechanical systems\cite{Latmiral2016}, cantilever systems\cite{Katz2007, Antonio2012, Badzey2004, Venstra2013} and Bose-Einstein condenstates\cite{Bertoldi2010, Makhalov2015}. This method can be used to estimate the magnitude of any kind of non-linearity with respect to displacement, as these affect the shape of the phase space distribution when a squeezing pulse is applied and the subsequent time evolution. In this way this method could be utilized to sense and quantify forces on the system. 

A similar, more sophisticated, method to match these two would be to use Sequential Monte Carlo (SMC) methods to estimate the non-linearity as it can optimize the value of the non-linearity $\xi$ as it simulates the system to best match the measurement record, however, this is much more computationally intensive.

\prlsec{Acknowledgements}
We would like to thank C. Timberlake for comments on the manuscript as well as M. Toro\v{s}, T. Georgescu and M. Rashid for discussions. We also wish to thank the Leverhulme Trust and the Foundational Questions Institute (FQXi) for funding. A. Setter is supported by the Engineering and Physical Sciences Research Council (EPSRC) under Centre for Doctoral Training grant EP/L015382/1. We also acknowledge support from EU FET project TEQ (grant agreement 766900). In addition the authors acknowledge the use of the IRIDIS High Performance Computing Facility, and associated support services at the University of Southampton. All data supporting this study are openly available from the University of Southampton repository at https://doi.org/10.5258/SOTON/D0967. The code used to analyse the data is openly available at https://doi.org/10.5281/zenodo.1042526.

\bibliographystyle{apsrev4-1}
\bibliography{library.bib} 

\end{document}